\documentclass[reprint,showpacs,preprintnumbers,aps,pre]{revtex4-1}
\usepackage{graphicx}
\usepackage{amssymb,amsmath}

\begin{document}
\title{Bounds and phase diagram of efficiency at maximum power for tight-coupling molecular motors}
\author{Z. C. Tu}\email{tuzc@bnu.edu.cn}
\affiliation{Department of Physics, Beijing Normal University, Beijing 100875, China}

\begin{abstract}The efficiency at maximum power (EMP) for tight-coupling molecular motors is investigated within the framework of irreversible thermodynamics. It is found that the EMP depends merely on the constitutive relation between the thermodynamic current and force. The motors are classified into four generic types (linear, superlinear, sublinear, and mixed types) according to the characteristics of the constitutive relation, and then the corresponding ranges of the EMP for these four types of molecular motors are obtained. The exact bounds of the EMP are derived and expressed as the explicit functions of the free energy released by the fuel in each motor step. A phase diagram is constructed which clearly shows how the region where the parameters (the load distribution factor and the free energy released by the fuel in each motor step) are located can determine whether the value of the EMP is larger or smaller than 1/2. This phase diagram reveals that motors using ATP as fuel under physiological conditions can work at maximum power with higher efficiency ($>1/2$) for a small load distribution factor ($<0.1$).
\end{abstract}
\preprint{Eur. Phys. J. E 36, 11 (2013)}
\maketitle

\section{Introduction}
Molecular motors are special proteins which are in charge of long-distance transports in living cells \cite{Howard}. The energetics of motors have attracted much attention from physicists since the 1990s \cite{Juelicher97,Reimann02,Haenggi09,Parrondo02,Astumian99,Oster02,Suzuki03}. It is intuitive that the operations of molecular motors should follow economic principles of energy utilization as molecular motors are the products of long-term evolution and natural selection.

One of the possible economic principles might be that molecular motors work at the region of maximum power. To clarify this principle, Schmiedl and Seifert \cite{Schmiedl2008mt} optimized the power of molecular motors based on a minimal model and found that the efficiency at maximum power (EMP) for tight-coupling molecular motors was 1/2 when the motors worked in the linear non-equilibrium region where the free energy released by the fuel in each motor step was very small. They also derived that the value of the EMP could be larger or smaller than 1/2 when the motors operated in the region far from equilibrium \cite{Schmiedl2008mt}. These results have been well confirmed and developed by other model systems~\cite{Seifert11,Seifert12rev,Golubeva12,Schulman12,Esposito12,VdBroeck12}, which can be regarded as the counterparts of the EMP for heat engines \cite{Curzon1975,chenyan,vdbrk2005,Schmiedl2008,Esposito2009,MEPRL2010,GaveauPRL2010,antonioPRL07,WTPRE2012,WangHeWu12}.
These investigations reveal that the EMP depends on the load distribution factor (i.e., the position of the transition state) and the free energy released per fuel molecule (for example, the hydrolysis energy of ATP). However, it is still
unclear under what conditions these parameters determine whether the value of the EMP is larger or smaller than 1/2---the universal EMP for tight-coupling motors under the approximation of linear response \cite{Schmiedl2008mt,Seifert11,VdBroeck12}. In other words, we lack good criteria by which to judge whether the performance of tight-coupling motors will increase or decrease when the motors move into the non-equilibrium regime. Therefore, it is necessary to construct a phase diagram which definitively tells us whether the value of the EMP is larger or smaller than 1/2 according to the region where these parameters are located.

Although the EMP can be larger or smaller than 1/2, it should be bounded between 0 and 1. A natural and significant issue is to investigate the existence of more accurate bounds. Recently, Van den Broeck \emph{et~al.}~\cite{VdBroeck12} derived the upper and lower bounds of the EMP for tight-coupling molecular motors, and found that these two bounds could be reached when the transition state was close to the reactant state or the product state, respectively. They expressed the bounds of the EMP as a function of the thermodynamic force which depends on both the free energy released by the fuel in each motor step and the external load on the motors. However, the released free energy and the external load are not independent at maximum power. It is useful to express the bounds as the explicit functions of a single variable---the free energy released by the fuel in each motor step.

Our main goal in this work is to address the two key issues mentioned above. We will analytically derive the lower and upper bounds of the EMP for tight-coupling motors, and then construct a phase diagram to show how the parameters can determine whether the value of the EMP is larger or smaller than 1/2.

\section{Minimal model\label{sec-model}}

\begin{figure}[!htp]
\begin{center}\includegraphics[width=6cm]{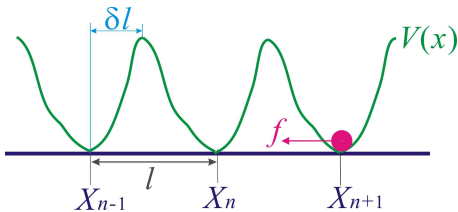}\end{center}
\caption{(Color online) Scheme of the energy landscape $V(x)$. $l$ and $\delta$ represent the step size and the load distribution factor.}\label{graph1}
\end{figure}

We consider a translational molecular motor with equivalent, discrete sites $X_n$
$(n=0,\pm 1, \pm 2, \cdots)$ with distance $l$ between the nearest neighbor sites~\cite{Schmiedl2008mt}. As is schematically shown in Fig.~\ref{graph1}, the motor will consume one fuel molecule and move against the external load $f$ in each step. That is, the chemical step and
the mechanical step are tightly coupled. This process can be regarded as a chemical reaction
\begin{equation}\mathrm{Fuel} + X_n \underset{\omega _{-}}{\overset{\omega _{+}}{\rightleftharpoons }} X_{n+1} + \mathrm{Products},\end{equation}
where \begin{equation}\omega _{+}=k_{+}\mathrm{e}^{-\delta fl},\label{ratef}\end{equation}
and
\begin{equation}\omega _{-}=k_{-}\mathrm{e}^{(1-\delta)f l},\label{rateb}\end{equation} are the forward and backward rate constants, respectively. We have not explicitly written out the energy scale ($k_B T$) of thermal motion at the physiological temperature in the expressions for the rate constants. Thus $fl$, in fact, represents $fl/k_B T$ in the present work. $k_{+}$ ($k_{-}$) equates to the bare forward (backward) rate constant times the concentration of fuel (products). The load distribution
factor $0\leq\delta\leq 1$ indicates the position of the transition state, i.e., the distance between the transition state and the reactant state. The
extreme cases $\delta=0$ and 1 correspond to the situations where the external load merely influences the backward
or forward rate constants, respectively. Thermodynamic consistency~\cite{Schmiedl2008mt}
requires
\begin{equation}\omega _{+}/\omega _{-}=\mathrm{e}^{\Delta \mu-fl},\label{thermconsic}\end{equation}
where $\Delta \mu$ is called the reduced chemical potential, which is the ratio of the free energy released  by the fuel in each motor step to the energy scale $k_B T$.
Interestingly, this minimal model was also adopted in very recent work~\cite{Shuyaogen} to show that the length of the neck linker in wild-type kinesin motors might be optimized for the largest possible stepping velocity (i.e. the maximum power).

\section{Constitutive relation and EMP\label{sec-constit}}
Let us consider the steady state, where the entropy production rate may be expressed as
\begin{equation}\sigma=(\omega _{+}-\omega _{-})\ln(\omega _{+}/\omega _{-})\label{entprodrate}\end{equation}
where the prefactor has been omitted.
The net thermodynamic current is defined as
\begin{equation}J=\omega _{+}-\omega _{-}.\label{eq-flux0}\end{equation} Thus, the fundamental thermodynamic relation (\emph{entropy production rate} $=$ \emph{thermodynamic current} $\times$ \emph{thermodynamic force}) allows us to express the thermodynamic force as
\begin{equation}F=\ln(\omega _{+}/\omega _{-})=\Delta\mu - fl\label{thermforce}\end{equation}
where Eqs.~(\ref{thermconsic})-(\ref{eq-flux0}) have been used. By substituting Eqs.~(\ref{ratef}), (\ref{rateb}) and (\ref{thermforce}) into (\ref{eq-flux0}), the constitutive relation between the thermodynamic current and force can be expressed as
\begin{equation}J=k_{-}\mathrm{e}^{(1-\delta)\Delta\mu}\left[\mathrm{e}^{\delta F} -\mathrm{e}^{-(1-\delta) F}\right]\label{eq-flux}.\end{equation}

The energy input per unit time for the tight-coupling motor can be expressed as $G=(k_B T\Delta \mu) J$.
The power output is the difference between the energy input and the energy dissipation per unit time, which can be expressed as
\begin{equation}P=G-k_B T\sigma=k_B T(\Delta \mu-F)J.\label{eq-power}\end{equation}
Thus, the efficiency can be defined as
\begin{equation}\eta=P/G=1-F/\Delta \mu.\label{eq-efficy}\end{equation}
It is stressed that we adopt the traditional definition of efficiency in the present work rather than the Stokes efficiency, rectification efficiency or sustainable efficiency proposed recently (Refs.~\cite{Astumian99,Oster02,Suzuki03,GaveauPRL2010}).

Our discussions so far can essentially be regarded as a concrete representation of the general considerations for the EMP of molecular motors in Refs.~\cite{Seifert11,Seifert12rev,VdBroeck12}. Now, maximizing the power (\ref{eq-power}) with respect to the external load $f$, we obtain
\begin{equation}J/J^\prime+F=\Delta \mu,\label{eq-optp}\end{equation}
where $J^\prime$ represents the derivative of the current $J$ with respect to the thermodynamic force $F$. Substituting the above equation into Eq.~(\ref{eq-efficy}), we find that the EMP ($\eta_{\ast}$) satisfies the following relation:
\begin{equation}\eta_{\ast}=\frac{1}{1+J^\prime F/J}.\label{eq-emp1}\end{equation}
This is the first main result in this work which obviously reveals that the EMP depends merely on the constitutive relation between the current $J$ and the thermodynamic force $F$.

\section{Classification of motors and corresponding ranges of the EMP\label{sec-class}}
\begin{figure}[!htp]
\begin{center}\includegraphics[width=6cm]{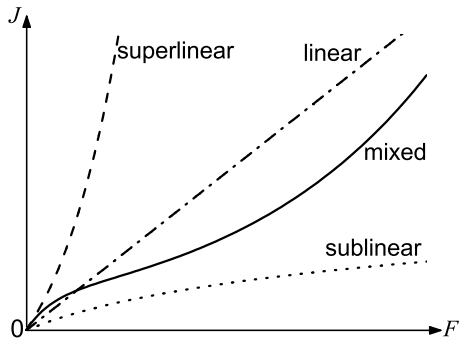}\end{center}
\caption{\label{fig-3type} Schematic diagram of four generic types of constitutive relations between the current $J$ and the thermodynamic force $F$.}\end{figure}

As we have done for heat engines in Ref.~\cite{Wangtu12}, we may classify motors into four generic types according to the characteristics of the constitutive relation. As shown in Fig.~\ref{fig-3type}, the motor is classified as linear when the constitutive relation is linear, i.e., $J$ increases uniformly with $F$. Similarly, a motor is superlinear (sublinear) if the rate of increase of $J$ is enhanced (reduced) by an increase in $F$ along the whole curve.
Additionally, the motor is of mixed type if the rate of increase of $J$ is enhanced by an increase in $F$ in some segments, but reduced in others. In fact, motors of this type may cross from sublinear to linear, and then to superlinear  with increasing $F$, or vice versa.

The linear, superlinear and sublinear types of constitutive relation can be mathematically expressed as $J=J^\prime F$, $J<J^\prime F$, and $J>J^\prime F$ for any finite $F$, respectively.
Considering $0<F<\Delta \mu$ and Eq.~(\ref{eq-emp1}), we can easily obtain
\begin{equation}\label{eq-bnd3type}
\left\{
                \begin{array}{ll}
                  \eta_{\ast}=1/2, & \hbox{linear tpye;} \\
                  0<\eta_{\ast}<1/2, &\hbox{superlinear type;} \\
                  1/2<\eta_{\ast}<1, & \hbox{sublinear type}.
                \end{array}
              \right.
\end{equation}
The constitutive relation of mixed type is more complicated than the above three types, which can be mathematically expressed as $J>J^\prime F$
in some segment (sublinear type) while $J<J^\prime F$ in another (superlinear type), and $J=J^\prime F$ at the demarcation point (linear type) between both segments. Correspondingly, the EMP can be larger or smaller than 1/2, or equal to 1/2 for a motor of mixed type, depending on the region in which $F$ is located. That is, the EMP for a motor of mixed type satisfies $0<\eta_{\ast}<1$. This fact and expression (\ref{eq-bnd3type}) form the second main result in the present work.

Now we will discuss three specific examples as follows. The first example is the situation that $\Delta\mu \ll 1$ so that $F\ll 1$. In this case, $J$ is approximately a linear function of $F$. According to Eq.~(\ref{eq-bnd3type}), we derive $\eta_{\ast}\rightarrow 1/2$ for $\Delta\mu \ll 1$, which confirms the previous result obtained in Refs.~\cite{Schmiedl2008mt,Seifert11,Seifert12rev,Golubeva12,Schulman12,Esposito12,VdBroeck12}.

The second example is the extreme case of $\delta =1$. On the one hand, substituting $\delta =1$ into Eq.~(\ref{eq-flux}), we derive the current $J=k_{-}(\mathrm{e}^F -1)$
which displays superlinear behavior ($J<J^\prime F$) for any finite $F$. Thus Eq.~(\ref{eq-bnd3type}) implies $0<\eta_{\ast}<1/2$. On the other hand, substituting the current into Eq.~(\ref{eq-optp}), we can obtain the optimized $F_{\ast}=\Delta\mu  -1 +\mathcal{W}(\mathrm{e}^{1-\Delta\mu})$ where $\mathcal{W}(.)$ is Lambert's W function~\cite{Corless,Yanshiw}. Substituting the optimized $F_{\ast}$ into Eq.~(\ref{eq-efficy}), we finally derive \begin{equation}\eta_{1}\equiv\eta_{\ast}(\delta =1)=\frac{1-\mathcal{W}(\mathrm{e}^{1-\Delta\mu})}{\Delta\mu}.\label{eq-empdet1}\end{equation}
This analytic expression is plotted in Fig.~\ref{fig-solved}, which reveals that the EMP indeed satisfies $0<\eta_{\ast}<1/2$.

The third example is the other extreme case of $\delta =0$. On the one hand, substituting $\delta =0$ into Eq.~(\ref{eq-flux}), we derive the current $J=k_{-}\mathrm{e}^{\Delta\mu }(1-\mathrm{e}^{-F})$ which displays sublinear behavior ($J>J^\prime F$) for any finite $F$. Thus Eq.~(\ref{eq-bnd3type}) implies $1/2<\eta_{\ast}<1$. On the other hand, substituting the current into Eq.~(\ref{eq-optp}), we can obtain the optimized $F_{\ast}=\Delta\mu  +1 -\mathcal{W}(\mathrm{e}^{1+\Delta\mu})$. Substituting the optimized $F_{\ast}$ into Eq.~(\ref{eq-efficy}), we finally derive \begin{equation}\eta_{0}\equiv\eta_{\ast}(\delta =0)=\frac{\mathcal{W}(\mathrm{e}^{1+\Delta\mu})-1}{\Delta\mu}.\label{eq-empdet0}\end{equation}
This analytic expression is plotted in Fig.~\ref{fig-solved}, which reveals that the EMP indeed satisfies $1/2<\eta_{\ast}<1$.

In the general case, we can prove from Eq.~(\ref{eq-flux}) that the motor is superlinear ($J<J^\prime F$) for $1/2\leq\delta<1$. Thus, the EMP should be smaller than 1/2 for $1/2\leq\delta<1$. On the other hand, it is not hard to find that the motor is of mixed type for $0<\delta<1/2$ through simple calculations from Eq.~(\ref{eq-flux}), and that the corresponding constitutive relation displays behavior similar to the solid curve in Fig.~\ref{fig-3type}. Then, the EMP can be larger or smaller than 1/2 for different values of $\Delta \mu$. A straightforward problem is to investigate the exact bounds of the EMP for a given value of $\Delta \mu$.

\section{Exact bounds of EMP for given $\Delta \mu$\label{sec-bdpt}}
Substituting Eq.~(\ref{eq-flux}) into Eq.~(\ref{eq-optp}), we derive
\begin{equation}\delta = \frac{1}{\Delta\mu-F}-\frac{1}{\mathrm{e}^F-1}.\label{eq-gendelt}\end{equation}
We cannot achieve the analytic solution to this equation if $\delta\neq 0$ or $1$. However, we readily see that $\delta$ is a monotonically increasing function of $F$ for a given $\Delta\mu$ because we find $\partial \delta/\partial F>0$ from the above equation. Equivalently speaking, $F$ is also a monotonically increasing function of $\delta$ for a given $\Delta\mu$. Additionally, Eq.~(\ref{eq-efficy}) implies that the efficiency is a monotonically decreasing function of $F$. Thus $\eta_{\ast}$ should be a monotonically decreasing function of $\delta$ for a given $\Delta\mu$. It follows that $\eta_{\ast}$ is bounded between $\eta_{1}$ and $\eta_{0}$, that is
\begin{equation}\frac{1-\mathcal{W}(\mathrm{e}^{1-\Delta\mu})}{\Delta\mu}\leq\eta_{\ast}\leq\frac{\mathcal{W}(\mathrm{e}^{1+\Delta\mu})-1}{\Delta\mu},\label{empbound}\end{equation}
which is the third main result in this work.
We note that Van den Broeck \emph{et al.} also derived the bounds of $\eta_{\ast}$ in Ref.~\cite{VdBroeck12}. However, they expressed the bounds as functions of $F$, which depends on both $\Delta\mu$ and $f$. Here, we express the bounds as the explicit functions of a single variable, $\Delta\mu$, which is an independent input parameter for the motors.

\begin{figure}[!htp]
\begin{center}\includegraphics[width=6cm]{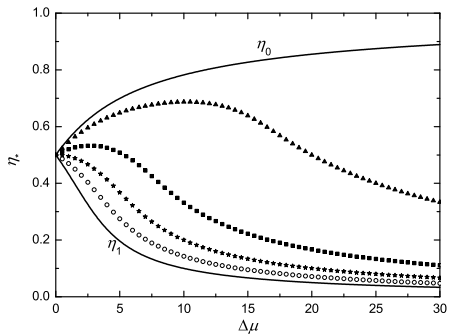}\end{center}
\caption{\label{fig-solved} EMP of molecular motors. $\eta_1$ and $\eta_0$ represent the EMP for $\delta =1$ and 0, which can be analytically expressed as Eqs.~(\ref{eq-empdet1}) and (\ref{eq-empdet0}), respectively. The triangles ($\blacktriangle$), squares ($\blacksquare$), stars ($\bigstar$) and circles ($\circ$) correspond to the EMP for $\delta =0.1$, 0.3, 0.5 and 0.7, respectively.}\end{figure}

We also find numerical solutions to Eq.~(\ref{eq-gendelt}) for $\delta=0.1$, 0.3, 0.5 and 0.7 with various values of $\Delta\mu$. Then the numerical relations between $\eta_{\ast}$ and $\Delta\mu$ are achieved from Eq.~(\ref{eq-efficy}) and shown in Fig.~\ref{fig-solved} as triangles, squares, stars and circles for $\delta=0.1$, 0.3, 0.5 and 0.7, respectively. Obviously, all numerical data are indeed located in the region bounded between $\eta_1$ and $\eta_0$, which is consistent with Inequality (\ref{empbound}). In addition, $\eta_\ast$ is always smaller than 1/2 for $\delta=0.5$ and $0.7$ while it can be either larger or smaller than 1/2 for $\delta=0.1$ and 0.3. This fact is consistent with our above discussions because the constitutive relations corresponding to $\delta=0.5$ and $0.7$ are of superlinear type while those corresponding to $\delta=0.1$ and 0.3 are of mixed type.

\section{Phase diagram\label{sec-phdg}}

There are two parameters in the present model: one is the load distribution factor $\delta$; another is the reduced chemical potential, $\Delta\mu$. We will consider how these two parameters can determine whether $\eta_{\ast}$ is larger or smaller than $1/2$. Since we have proved $\eta_{\ast}\rightarrow 1/2$ in the linear case of $\Delta\mu\rightarrow 0$, we only discuss the case of $\Delta\mu>0$ in the following contents.
To address this problem, we first find the condition (the relation between $\delta$ and $\Delta\mu$) to make $\eta_{\ast}=1/2$. It follows that $J^\prime F/J =1$ and $F=\Delta\mu/2$ from Eqs.~(\ref{eq-optp}) and (\ref{eq-emp1}). Combining Eq.~(\ref{eq-flux}), we derive
\begin{equation}\delta =\frac{2}{\Delta\mu}-\frac{1}{\mathrm{e}^{\Delta\mu/2}-1}.\label{phasebd}\end{equation}
That is, when $\delta$ and $\Delta\mu$ satisfy the above relation, $\eta_{\ast}$ should always be 1/2.
Since we have proved that the efficiency is a monotonically decreasing function of $\delta$ for a given $\Delta\mu$ in the above discussion, we can readily deduce that $\eta_{\ast}>1/2$ if $\delta < 2/{\Delta\mu}-1/(\mathrm{e}^{\Delta\mu/2}-1)$ and vice versa. This is the fourth main result in the present work.

\begin{figure}[!htp]
\begin{center}\includegraphics[width=6.5cm]{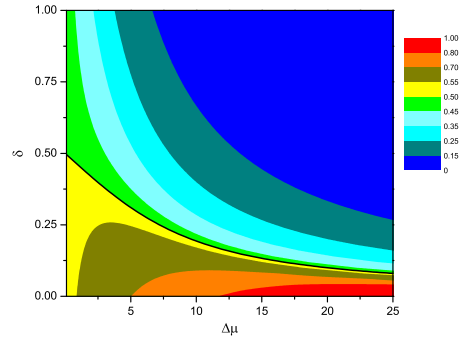}\end{center}
\caption{(Color online) Phase diagram. The contour plot displays the EMP for various values of the parameter pair ($\Delta\mu,\delta$). The thick curve is the phase boundary which separates the parameter plane into two regions. The value of the EMP is smaller than 1/2 when the parameter pair ($\Delta\mu,\delta$) takes values in the region above the boundary, and larger than 1/2 below the boundary.\label{fig-phdiag}}\end{figure}

We draw the phase diagram in Fig.~\ref{fig-phdiag} according to the above analysis. The phase boundary is described by Eq.~(\ref{phasebd}). The value of the EMP is smaller than 1/2 when the parameter pair ($\Delta\mu,\delta$) takes values in the region above the boundary, and larger than 1/2 below the boundary. We also calculate the numerical results of $\eta_{\ast}$ for various values of $\delta$ and $\Delta\mu$ by using Eqs.~(\ref{eq-efficy}) and (\ref{eq-gendelt}), and then draw a contour plot in Fig.~\ref{fig-phdiag} which exquisitely displays how the EMP of tight-coupling motors are determined by the values of the parameter pair ($\Delta\mu,\delta$). Obviously, these numerical results indeed support our above theoretical predictions. This phase diagram implies that the non-equilibrium effect will increase the performance of tight-coupling motors when $\delta < 2/{\Delta\mu}-1/(\mathrm{e}^{\Delta\mu/2}-1)$, but decrease the performance when $\delta > 2/{\Delta\mu}-1/(\mathrm{e}^{\Delta\mu/2}-1)$.

In Ref.~\cite{Schmiedl2008mt}, Schmiedl and Seifert argued that long-term evolution and natural selection might shape the ability of motors to work at maximum power with higher efficiency.
This insight can be easily understood with the aid of the phase diagram according to which the parameter pair ($\Delta\mu,\delta$) should be located in the region below the boundary. For motors using ATP as fuel, $\Delta\mu \approx 20$ under physiological conditions, we calculate $\delta < 2/{\Delta\mu}-1/(\mathrm{e}^{\Delta\mu/2}-1) \approx 0.1$, which is in good agreement with several experiments on kinesin and myosin motors where a small $\delta$ ($<0.1$) was observed for the main motor step \cite{Clemen05,Fisher01,Kolomeisky03,Gebhardt06}. This fact suggests a more specific economic principle of energy utility for molecular motors: long-term evolution and natural selection shape the good performance of molecular motors so that they can operate at maximum power with efficiency larger than 1/2. A possible interpretation is as follows: If the efficiency of some motors is smaller than 1/2, then the energy dissipation is larger than the useful energy. That is, most of the input energy is wasted by the motors. Such motors might be unfavorable for natural selection, therefore they will easily become extinct under long-term evolution.

\section{Conclusion and discussion\label{sec-conclud}}

In the above discussions, we investigate the EMP for tight-coupling motors and find that the EMP depends merely on the constitutive relation between the thermodynamic current and force. We derive the bounds of the EMP [Inequality (\ref{empbound})] and construct a phase diagram [Fig.~\ref{fig-phdiag}], from which we can easily see under which conditions the EMP is larger or smaller than 1/2. We infer $\delta < 0.1$ for motors using ATP as fuel under physiological conditions, which is consistent with several experimental observations on kinesin and myosin motors. Finally, we would like to clarify the following two points which have not been emphasized in the above discussions.

i) The analysis in the present work might be extended to discuss the bounds of the EMP for information machines or the Feynman ratchet investigated in recent literature~\cite{Bauer12,Schaller-Esp12,Tu2008}. At a glance, the EMP of the Feynman ratchet derived in our previous work \cite{Tu2008} might in fact be the lower bound because it corresponds to the case of $\delta=1$. However, in the recent work by Van den Broeck and Lindenberg~\cite{vdb-lin2012}, it was found that the EMP for classical particle transport and the EMP of the Feynman ratchet share the same expression, which seems to imply that the EMP of the Feynman ratchet should be independent of $\delta$. It is indeed not hard to verify this point based on the discussions in the present work.

ii) All discussions in the present work are merely focused on a minimal model with discrete sites where the internal states are neglected. While the main prediction $\delta< 0.1$ of this minimal model is in good agreement with most of experimental observations, a transition state in the range $\delta \simeq 0.3-0.65$ has also been extracted on the basis of a six-state model~\cite{Lipowsky07}. It is necessary to investigate how the model with internal states influences the bounds and phase diagram in this work.\\

The author acknowledges the financial support of the National Natural Science Foundation of China (grant NO. 11075015) and the Fundamental Research Funds for the Central Universities. He thanks Shiwei Yan for drawing his attention to Lambert's W function. The author also thanks Yang Wang and Shiqi Sheng for carefully proofreading the present paper.

\end{document}